\documentclass[10pt, a4paper]{article}
\usepackage{xcolor}
\usepackage{listings}
\usepackage{subcaption}
\usepackage{graphicx}
\usepackage{tabularx}
\usepackage{amsfonts}
\usepackage{amsmath}
\usepackage[colorinlistoftodos,textsize=small]{todonotes}

\usepackage{lrec2026} 

\title{Do LLMs Ask the Right Questions?  Evaluating GPT-Generated Surveys as Instruments for Measuring Social Attitudes}

\name{Tina Behzad, Wenbo Li, Reuben Kline, Klaus Mueller} 

\address{
         Stony Brook University \\
         \{tbehzad, mueller\}@cs.stonybrook.edu, \{wenbo.li.1, reuben.kline\}@stonybrook.edu
}

\abstract{
Understanding human beliefs and social attitudes often relies on carefully designed survey instruments. Recent work has suggested that large language models (LLMs) could automate parts of this process by generating surveys at scale, raising questions about the comparability of such instruments to literature-grounded, human-designed surveys. We present a controlled empirical comparison between GPT-generated surveys and established survey baselines across three social domains: climate change, immigration, and diversity, equity, and inclusion (DEI). GPT-generated surveys were produced using a fixed prompting framework enforcing a 3×3 structure over beliefs, perceptions, and behaviors, while human baselines were assembled from validated instruments to match survey length and construct coverage. We collected responses from U.S.-based participants, who completed both survey types, allowing direct within-subject comparison. We analyze differences in response distributions, clustering behavior, and alignment with self-identified stances. Our results show that GPT-generated surveys capture the same dominant attitudinal divisions as human-designed instruments, while exhibiting differences in the resolution of belief structure and group separation. These findings suggest that LLM-generated surveys are suited for exploratory and large-scale analyses, and can be used to complement expert-designed instruments.
 \\ \newline  }

\begin{document}

\maketitleabstract
\thanks{Originally published at SoCon/NLPSI @ LREC 2026, pp. 48--66.}
\section{Introduction}
Surveys are a central tool for studying human beliefs, values, and social attitudes across disciplines such as psychology, political science, and sociology. Carefully designed survey instruments are not merely collections of questions, but theory-informed measurement tools that encode assumptions about which constructs matter, how they should be operationalized, and how responses should be interpreted \cite{dillman2014internet, boynton2004selecting,synodinos2003art}. As a result, survey design choices such as question framing, construct coverage, and scale structure play a critical role in shaping the conclusions drawn about public opinion and social behavior \cite{knauper1997question,lenzner2012effects}.

Recent advances in large language models (LLMs) have raised the possibility of automating or augmenting survey design. Prior work has shown that LLMs can generate fluent survey questions \cite{maiorino2023application}, simulate respondent behavior \cite{argyle2023out,bisbee2024synthetic}, and scale social data collection \cite{wuttke2025ai}, facilitating rapid pilot studies, iterative instrument refinement, and large-scale processing of unstructured responses, tasks that are typically resource-intensive and slow in survey workflows.

In this work, we present a systematic comparison between GPT-generated surveys and human-designed surveys as a baseline for measuring beliefs and stances on social topics. Such comparisons are essential for determining whether LLMs can be responsibly used as tools in questionnaire design without compromising measurement validity or analytic conclusions.
We focus on three timely and socially consequential domains: climate change, immigration, and diversity, equity, and inclusion (DEI).

We evaluate whether surveys generated using a fixed, structured prompt template produce coherent and reliable instruments across topics and elicit response patterns comparable to those obtained using established instruments. Rather than assessing surface-level similarities in wording or face validity, we examine how the two survey types differ in their ability to capture belief variation, support meaningful clustering of respondents, and align with self-identified stances. 

Our contributions are threefold. First, we introduce a framework for generating surveys using LLMs that explicitly controls construct coverage and survey length, enabling direct comparison with human-designed instruments. Second, we conduct an empirical evaluation of LLM-generated and human-designed surveys across multiple social domains, using responses collected online from 150 U.S.-based participants per topic. Third, we analyze where and how GPT-generated surveys diverge from human baselines in ways that matter for interpreting social beliefs, highlighting both their potential and their limitations as measurement tools.

Taken together, our findings contribute to ongoing discussions at the intersection of NLP and psychology by clarifying when LLM-generated surveys may be useful and where caution is warranted. More broadly, this work underscores the importance of treating survey generation not as a purely linguistic task, but as a measurement problem with substantive implications for how social attitudes are modeled and understood.

\section{Related Work}
This section reviews foundational approaches to survey design, emerging work on LLM-based survey generation, and common approaches to quantify and evaluate survey quality.
\subsection{Measuring Social Attitudes}
Survey instruments are central to measuring beliefs and social attitudes, but their validity depends on how abstract constructs are operationalized into items, response scales, and questionnaire context \cite{groves2011survey}. Classic work in survey methodology demonstrates that seemingly small design choices such as question wording, ordering, and framing, can systematically shift responses and even change the meaning respondents assign to a question \cite{schuman1996questions}. More broadly, survey methodology emphasizes that surveys are subject to multiple sources of error (e.g., comprehension, recall, judgment, and response mapping), and that careful questionnaire construction is necessary for producing interpretable and reliable measurements\cite{groves2011survey}. 

Survey instruments aimed at measuring social attitudes typically begin with the definition of an underlying construct space that specifies which dimensions of belief are to be measured and how they relate to one another. Rather than treating attitudes as unitary variables, a large body of work in psychology and social science conceptualizes attitudes as multidimensional, often distinguishing between cognitive beliefs, affective evaluations, and behavioral or policy-oriented intentions \cite{ajzen1991theory, eagly1993psychology}. This tripartite view has informed the design of many survey instruments, particularly in domains where attitudes are complex, value-laden, and socially contested.

Operationalizing construct spaces into survey items has been approached through multiple methodological strategies. Some surveys rely on multi-item, construct-based scales, in which parallel items are designed to probe distinct facets of an underlying construct \cite{cronbach1955construct}, while others adopt segmentation-oriented designs that prioritize a small number of high-signal questions to differentiate respondent groups rather than exhaustively measure all dimensions \cite{chryst2018global}. In applied and large-scale survey settings, these design choices are typically shaped by a combination of theoretical grounding, prior empirical usage, and practical constraints such as respondent burden and survey length \cite{groves2011survey}. Across approaches, a common principle is the use of structured and balanced item layouts that explicitly sample across dimensions to support interpretability and comparability, facilitating downstream analyses such as clustering and segmentation \cite{saris2014design}.
\subsection{LLM-Based Survey Generation}
The emergence of LLMs has created new opportunities across the survey lifecycle, spanning instrument design, administration, and response analysis. Prior work has explored LLMs as generators of synthetic survey responses \cite{argyle2023out,hamalainen2023evaluating}, as tools for survey question selection, editing, and drafting \cite{rothschild2024opportunities}, and as aids for annotating and analyzing open-ended responses \cite{valenzuela2025using,jansen2023employing}. 

Early work on question generation demonstrated that LLMs can support researchers in editing and paraphrasing survey questions, as well as in redesigning response options for multiple-choice items \cite{rothschild2024opportunities}. More recently, a growing body of work has shifted toward using LLMs to generate entire questionnaires. For example, \citet{adhikari2025exploring} study how LLMs can be prompted to produce questionnaire items that adhere to specified formats, constructs, and design constraints, highlighting their potential for rapid survey creation and customization. Related studies have examined the use of LLM-generated surveys in applied settings such as communication research \cite{valenzuela2025using}, market research \cite{keane2025using}, and personality measurement \cite{oeljeklaus2025comparing}.

In contrast to prior work that emphasizes topic-specific applications or task-tailored prompting, we investigate whether LLMs can be used in a more generalizable manner by employing a single, universal prompt to generate questionnaires across diverse social topics, and assess how reliably these instruments compare to expert-designed surveys.

\subsection{Evaluating Survey Quality}
A common approach to evaluating survey quality in the social sciences is to examine whether responses support the discovery of meaningful latent structure and are predictive of downstream intentions, behaviors, or policy-relevant outcomes. A substantial literature demonstrates that clustering and latent-class analyses of survey data can recover meaningful population segments that differ systematically in behaviors, participation patterns, and policy preferences, motivating their use as analytic tools rather than purely descriptive techniques. For example, climate-opinion research has shown that audience segments derived from survey responses capture heterogeneity in engagement behaviors and policy support, enabling prediction and targeted communication strategies \cite{chryst2018global,maibach2011identifying,leiserowitz2021global}. Similarly, political science research has used latent class and profile models to identify participation and voting types that predict turnout and modes of civic engagement more effectively than item-level analyses \cite{greaves2015profiling,johann2020channels, oser2022protest}. More broadly, foundational work on attitude–behavior relationships establishes that survey-measured latent constructs are meaningful precisely because they support the prediction of intentions and behaviors across domains \cite{ajzen1991theory,mceachan2016meta}. 
Within this context, recent work on LLM-generated surveys has primarily evaluated generated instruments through comparisons with expert-designed questionnaires, often using psychometric criteria. Studies have assessed whether LLM-generated items reproduce the factor structure, internal consistency, and convergent validity of established scales, treating human designed surveys as reference measurement models \cite{adhikari2025exploring,terry2025artificial, oeljeklaus2025comparing}.
While these evaluations are typically tied to study-specific measures and predefined scales, we adopt similar principles within a more general evaluation framework. In particular, we extend assessment beyond construct replication to examine latent structure, respondent segmentation, stability, and predictive utility, thereby evaluating the robustness and analytic usefulness of the induced belief structure across domains.

\label{sec:append-how-prod}

\section{Methodology}
This section describes the experimental design and analysis procedures used in this study. 
We employ a comparative setup in which GPT-generated surveys are evaluated alongside 
literature-grounded, human-designed surveys targeting the same social topics. 
The study proceeds in four stages:

\begin{enumerate}
    \item \textbf{Survey Generation} (Section~\ref{sec:gpt-generation}): 
    GPT-generated surveys were produced using a controlled prompting framework 
    specifying construct structure, formatting constraints, and generation parameters.
    
    \item \textbf{Human Baseline Construction} (Section~\ref{sec:human-baseline}): 
    Literature-grounded survey instruments were assembled from established sources 
    to match the length and conceptual coverage of the GPT-generated surveys.
    
    \item \textbf{Response Collection} (Section~\ref{sec:response-collection}): 
    U.S.-based participants completed both survey types under randomized ordering, 
    along with additional survey-quality assessments.
    
    \item \textbf{Comparative Analysis} (Section~\ref{sec:metrics}): 
    We analyzed differences in survey structure and response patterns using 
    quantitative measures of construct coverage, clustering behavior, 
    and alignment with self-identified stances.
\end{enumerate}

For the purpose of this paper we focus on three timely and socially consequential domains: climate change, immigration, and diversity, equity, and inclusion (DEI)

These topics were selected for their broad scope and frequent presence in everyday discourse, making it likely that individuals across diverse backgrounds hold opinions on them, while also helping to reduce topic-specific confounds and enhance the generalizability of our findings.

\subsection{GPT-Generated Survey Construction} \label{sec:gpt-generation}
To structure the GPT-generated surveys, we adopt a 3×3 construct space that systematically varies both the substantive dimension of the topic and the type of attitude being elicited. This design choice is motivated by established survey methodology research, which emphasizes that social attitudes are multidimensional and are more reliably measured through structured sets of questions rather than single items \cite{tourangeau2000psychology, saris2014design,dillman2014internet}. By explicitly crossing topical facets with distinct attitudinal orientations, the 3×3 design promotes broader construct coverage while maintaining a controlled and interpretable structure. At the same time, this layout balances expressiveness with cognitive and methodological constraints, avoiding overly long or unstructured surveys while enabling consistent comparison across survey conditions.

GPT-generated surveys were created using a fixed prompting template applied consistently across topics. Survey generation was performed using the OpenAI API with GPT-4.1 model ( temperature of 0.7). A system prompt instructed the model to assume the role of an expert survey methodologist and social scientist, with the goal of producing concise, signal-rich surveys suitable for persona discovery and cluster analysis. The user prompt specified the target topic and imposed strict design constraints, including a 3×3 structure spanning beliefs, perceptions, and behaviors; neutral, non-leading wording; and a uniform 5-point Likert response format. Additional constraints required questions to be non-redundant, accessible to a general adult audience, and oriented toward maximizing differentiation between respondents when analyzed jointly. To support interpretability, the model was also instructed to include a brief parenthetical note describing the latent dimension probed by each question. The full prompt template alongside the final survey questions are provided in the supplementary material.

\subsection{Human Survey Baseline} \label{sec:human-baseline}
To enable meaningful comparison with the GPT-generated surveys, we constructed human survey baselines with a comparable scope and length. Because literature-grounded surveys on social topics often vary in granularity and focus, we combined items from multiple established survey instruments to end up with at least 8 questions. Item selection prioritized alignment with the same high-level constructs targeted in the GPT surveys, namely beliefs, perceptions, and behaviors, while preserving the original wording and intent of each question as used in prior work. This approach allows us to compare surveys of similar length and conceptual coverage, while maintaining fidelity to human-designed instruments grounded in existing social science research. We detail the construction of the human baseline survey for the climate change here. The procedures for the other two topics and the full survey questions are available in the supplementary material. Throughout the paper, we refer to this survey as the human baseline.

For the climate change domain, we constructed the human survey baseline by drawing on widely used and well-validated survey instruments in the literature. The core of the baseline survey is based on the Six Americas Short Survey (SASSY) \cite{chryst2018global}, which has been extensively used to segment public attitudes toward climate change and is supported by a large body of empirical reports and downstream analyses.

To complement this core instrument and expand construct coverage without duplicating question content, we incorporated the Single-Item Measure of Belief in Climate Change \cite{berger2025measuring}, which captures causal attribution and evaluative belief about climate change in a compact and validated form. We further added two non-redundant items drawn from the American National Election Studies (ANES) \cite{anes2020ts}, and one question from the climate change perceptions questionnaire \cite{poortinga2019climate} to capture perceptions of the current impacts of climate change, such as its perceived effects on severe weather events and temperature patterns in the United States.

\subsection{Response Collection} \label{sec:response-collection}
Participants were recruited via the Cloud Connect platform and were required to be at least 18 years old and residing in the United States. Data collection was conducted using Qualtrics, and the study protocol was approved by our university's Institutional Review Board (IRB). For each topic (climate change, immigration, and DEI), we recruited approximately 150 participants. Participants were compensated at an hourly rate of \$11; the average completion time was approximately 6 minutes. No personally identifiable information was collected, and detailed demographic statistics are provided in the supplementary material.

For each topic, participants first reviewed and signed an informed consent form. They were then asked to self-identify their stance using a predefined set of labels accompanied by brief descriptions (full label lists provided in the supplementary material). This self-identification served two purposes: (1) to support downstream analysis of how survey responses cluster relative to self-reported stances, and (2) to facilitate the collection of diverse viewpoints. To this end, once a threshold number of participants was reached for a given stance label, additional participants selecting that label were screened out to maintain a more balanced distribution of perspectives.

Each participant completed both the GPT-generated survey and the corresponding human survey baseline for the topic. Survey order was randomized to control for ordering effects. Following each survey, participants answered a short four-item questionnaire assessing perceived survey quality, including clarity, bias, relevance, and over-specificity.

\subsection{Evaluation Metrics} \label{sec:metrics}

To compare the quality of human baseline and GPT-generated surveys, we employ a set of complementary evaluation metrics that assess latent structure, robustness, diversity preservation, and behavioral relevance. All analyses that require clustering use $k$-means, a standard unsupervised clustering algorithm, applied to standardized survey responses.

\subsubsection{Clustering Quality and Stability}

We treat clustering quality as a central aspect of survey quality. In many social science applications, surveys are not used solely to analyze individual items in isolation, but to induce latent groups or typologies of respondents that summarize heterogeneity in beliefs, attitudes, or preferences. Such typological analyses are foundational in fields including political science, sociology, and psychology, where survey responses are routinely clustered to identify ideological blocs, belief systems, or behavioral profiles \cite{gross2012mixed,ahlquist2012model,thomas2019vegetarian}. From this perspective, a high-quality survey should support coherent, stable, and interpretable group structure, as deficiencies in clustering quality can propagate to downstream tasks such as belief modeling, behavioral prediction, and policy analysis.

We evaluate internal clustering quality using the Silhouette Score \cite{rousseeuw1987silhouettes} and Davies--Bouldin Index (DBI) \cite{4766909}, computed on cluster assignments obtained via unsupervised $k$-means clustering. The Silhouette Score assesses cluster separation at the level of individual respondents by comparing within-cluster cohesion to nearest-cluster separation. For a point $i$, it is defined as
\[
s(i) = \frac{b(i) - a(i)}{\max\{a(i), b(i)\}},
\]
where $a(i)$ is the average distance from $i$ to all other points in the same cluster, and $b(i)$ is the minimum average distance from $i$ to points in any other cluster. The overall Silhouette Score is computed as the mean of $s(i)$ across all $N$ points.
The score ranges from $-1$ to $1$, with higher values indicating stronger separation between clusters and values near zero suggesting overlapping structure.

We also report inertia, which measures overall within-cluster dispersion. Inertia is defined as the sum of squared distances between each point and the centroid $\mu_{c(i)}$ of its assigned cluster:
\[
\text{Inertia} = \sum_{i=1}^{N} \left\| x_i - \mu_{c(i)} \right\|^2,
\]
where $x_i$ denotes the feature vector for point $i$ and $c(i)$ denotes its cluster assignment. Lower inertia indicates more compact clusters, though it does not account for between-cluster separation.

The DBI evaluates cluster compactness relative to centroid separation and is defined as
\[
\mathrm{DBI} = \frac{1}{K} \sum_{k=1}^{K} \max_{l \neq k} 
\frac{\sigma_k + \sigma_l}{d(c_k, c_l)},
\]
where $\sigma_k$ denotes within-cluster scatter for cluster $k$ and $d(c_k, c_l)$ denotes the distance between cluster centroids. Lower DBI values indicate more compact and well-separated clusters, while higher values indicate poorer separation. Because these metrics emphasize complementary geometric properties, separation vs. compactness, agreement between them provides stronger evidence of coherent structure than reliance on a single index.

Since internal clustering metrics can be sensitive to sampling variation, we additionally assess \emph{stability under resampling} by repeatedly bootstrapping participants, reclustering with $k$-means, and computing the Adjusted Rand Index (ARI)~\cite{hubert1985comparing} between runs. 
The ARI between two partitions $U$ and $V$ is defined as
\[
\mathrm{ARI} = 
\frac{\mathrm{RI} - \mathbb{E}[\mathrm{RI}]}{\max(\mathrm{RI}) - \mathbb{E}[\mathrm{RI}]}
\]
where $\mathrm{RI}$ denotes the Rand Index, which measures pairwise agreement between two clusterings by evaluating all pairs of samples and counting the proportion assigned either to the same cluster in both partitions or to different clusters in both partitions. The ARI corrects this measure for chance agreement. ARI values range from $-1$ to $1$, with $0$ indicating chance-level agreement and $1$ indicating identical partitions. Higher ARI under resampling indicates more stable and robust latent structure.

\paragraph{External Structural Agreement.}
We evaluate the external validity of the induced cluster structure in two complementary ways. First, to assess whether different survey instruments for a single topic recover similar respondent groupings, we apply unsupervised $k$-means clustering independently to each survey and examine agreement between the resulting partitions using transition matrices. This analysis tests whether the two surveys capture a shared underlying organization of beliefs.

Second, we assess alignment with self-identified persona labels, which participants selected at the beginning of the study to describe their own stance. We compare survey-induced cluster assignments to these self-reported personas using ARI \cite{hubert1985comparing} and Normalized Mutual Information (NMI) \cite{strehl2002cluster}. Let $C_1$ denote the clustering obtained from survey responses and $C_2$ denote the partition induced by participants’ self-reported persona labels. NMI is defined as
\[
\mathrm{NMI}(C_1, C_2) = \frac{2 I(C_1; C_2)}{H(C_1) + H(C_2)},
\]
where $I(C_1; C_2)$ measures the mutual information between the two partitions and $H(C_1)$ and $H(C_2)$ denote their respective entropies. NMI ranges from $0$ (no shared information) to $1$ (identical partitions). Moderate alignment suggests that the derived clusters capture meaningful distinctions while remaining non-redundant with persona categories.

\subsubsection{Predictive Validity}

A key criterion for evaluating survey quality is the extent to which survey responses capture information that is relevant for downstream behaviors and attitudes. In social science and behavioral research, surveys are routinely used not only to describe beliefs, but to predict consequential outcomes such as political participation, policy support, and social engagement~\cite{ajzen1991theory,fishbein2011predicting, pascal2016turnout,lazarsfeld1945prediction}. From this perspective, predictive performance provides a direct test of whether a survey encodes behaviorally meaningful signal.
We assess predictive validity by evaluating how well survey responses and survey-induced clusters predict downstream outcomes, including self-reported policy support, action likelihood, and social discussion. Cluster-based predictors use $k$-means assignments as discrete representations of latent structure. Predictive performance is measured using cross-validated $R^2$, defined as
\[
R^2 = 1 - \frac{\sum_i (y_i - \hat{y}_i)^2}{\sum_i (y_i - \bar{y})^2}.
\]
Comparing feature-based models with cluster-based models reveals how much predictive signal is preserved under unsupervised compression.

\section{Results}
Due to space constraints, we focus our empirical evaluation on the climate change surveys, which serve as a representative testbed for comparing human-designed and GPT-generated instruments. Results for the remaining two topics follow similar patterns unless otherwise noted and are reported in the supplementary material. All reported results are computed using the same set of respondents, each of whom completed both the human baseline survey and the GPT-generated survey on climate change.
We begin our analysis by examining mean responses to each survey question, stratified by self-identified climate personas. For the climate change topic, participants were asked to select one of the following labels: \textbf{Advocate for Avoiding Climate Alarmism}, \textbf{Climate Policy Skeptic}, \textbf{Neutral Observer of Climate Change}, \textbf{Advocate for Balanced Climate Action}, \textbf{Advocate for Proactive Climate Policy}, and \textbf{Climate Action Advocate}.

These personas were designed to capture ordered positions along a spectrum of climate beliefs, reflecting increasing levels of concern about climate change and stronger support for mitigation and policy intervention. Each label was accompanied by a brief description (see Supplementary Material), and wording was deliberately framed in neutral or positive terms to avoid discouraging participants from selecting any category.

\begin{figure}[t]
    \centering
    \includegraphics[width=\linewidth]{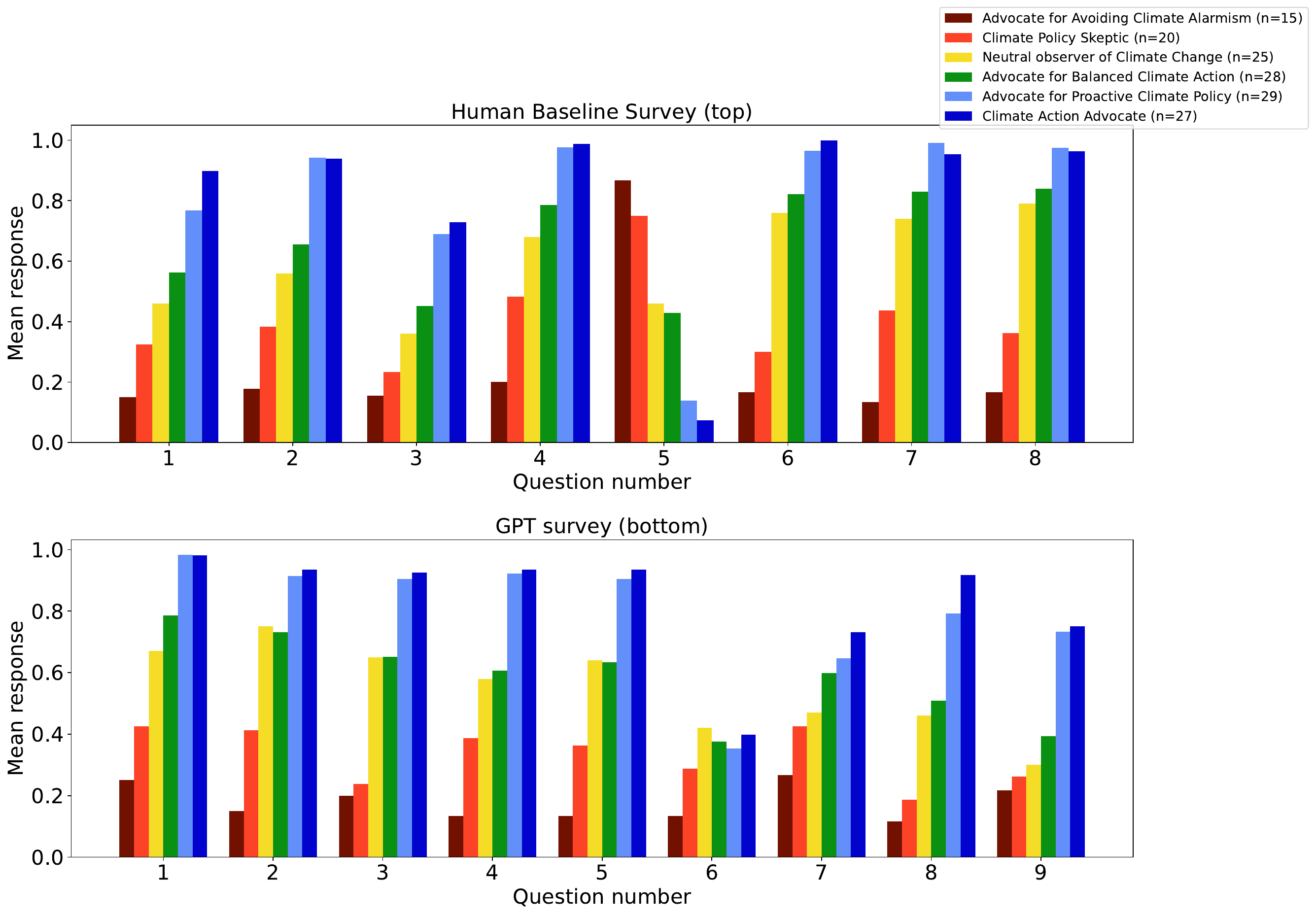}
    \caption{
    Climate change surveys: Normalized mean response per question by persona label.
    The top panel shows the human baseline survey and the bottom panel shows the GPT-generated survey.
    Bars are colored by self-identified climate persona, ordered from least to most climate-concerned.
    Across both surveys, mean responses vary monotonically along the persona spectrum, indicating an ordinally meaningful alignment between survey responses and underlying climate belief positions. 
    }
    \label{fig:means_by_persona}
\end{figure}

Figure~\ref{fig:means_by_persona} shows the mean response for each question in the human baseline survey (top) and the GPT-generated survey (bottom), with bars colored by persona. Across nearly all questions in both surveys, mean responses exhibit a clear and ordinally meaningful pattern: as one moves along the persona spectrum toward more climate-concerned identities, mean responses change monotonically in the expected direction. 

The monotonic structure indicates that both surveys capture a coherent underlying belief gradient rather than noisy variation. The human baseline shows more consistent spacing between personas, suggesting stable differentiation, whereas the GPT-generated survey exhibits attenuated separation across some questions, indicating partial merging of adjacent personas.

Overall, these results validate both the self-identified personas and the survey instruments, showing that persona labels correspond to systematic response patterns and that both surveys align with these distinctions.

\subsection{Clustering Quality and Stability}\label{sec:results:cluster}

\begin{figure}[t]
    \centering

    \begin{subfigure}{\linewidth}
        \centering
        \includegraphics[width=\linewidth]{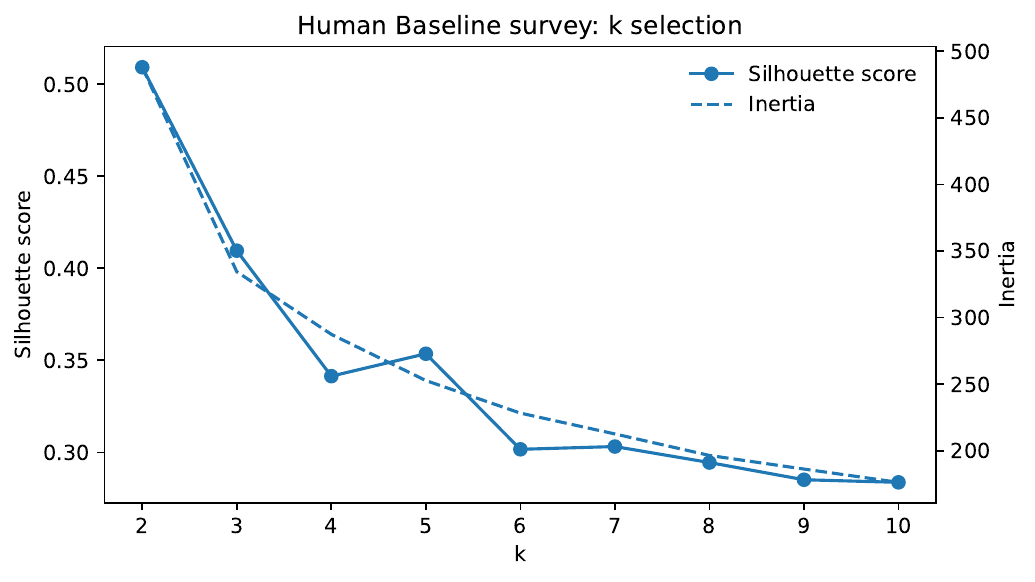}
        \label{fig:k_lit}
    \end{subfigure}

    \begin{subfigure}{\linewidth}
        \centering
        \includegraphics[width=\linewidth]{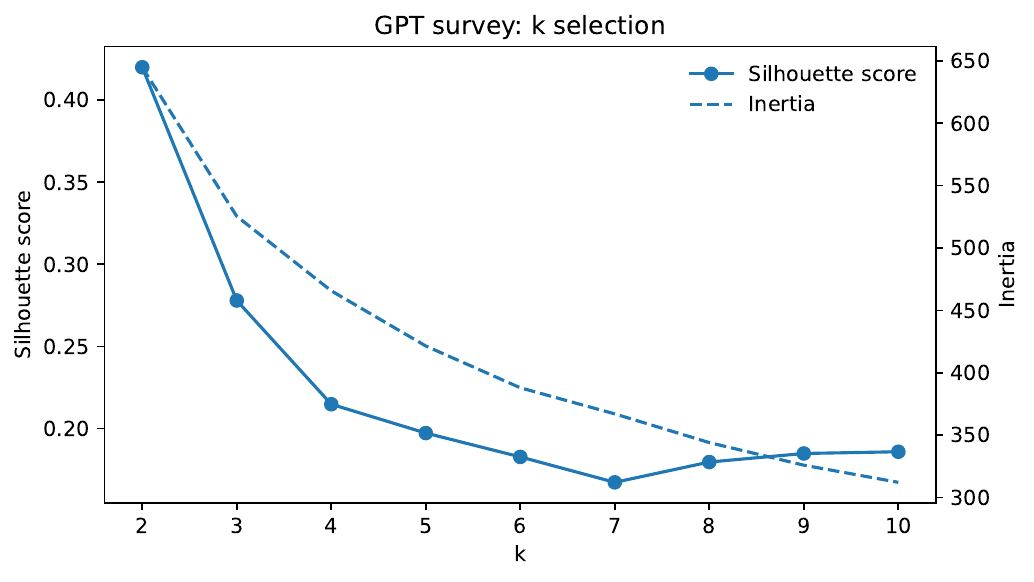}
        \label{fig:k_gpt}
    \end{subfigure}

    \caption{
    Selection of the number of clusters ($k$) using silhouette score and inertia for the human baseline(top) and GPT-generated(bottom) surveys. 
    }
    \label{fig:k_selection}
\end{figure}

We next examine the number of clusters supported by each survey using internal clustering metrics. Based on both the Silhouette Score and the DBI, the optimal number of clusters is $k=2$ for both the human baseline and the GPT-generated climate change surveys (as indicated by the Silhouette curve in Figure \ref{fig:k_selection}). Although we do not include DBI plots due to space constraints, both metrics consistently indicate that a two-cluster solution yields the most compact and well-separated structure for both surveys.

Although this may seem inconsistent with the six self-identified personas, we do not view it as contradictory. Instead, both surveys appear to capture a dominant, coarse-grained division in climate beliefs, consistent with prior work showing that attitudes often exhibit hierarchical structure \cite{goldberg2006doing}. Our initial analyses indicate that each of the two major clusters can be further subdivided into more nuanced subgroups using hierarchical or probabilistic approaches such as Latent Profile Analysis (LPA) \cite{masyn2013latent}. Exploring such multi-level structure is a promising direction for future work; in the remainder of this paper, we focus on the two-cluster solution.

We assess the robustness of the induced cluster structure using stability under resampling. For each survey, we perform $300$ bootstrap runs, each time sampling $80\%$ of respondents without replacement and reclustering using $k$-means with $k=2$. Stability is quantified using the ARI between cluster assignments across runs.

Both surveys exhibit high stability. The human baseline survey achieves a mean ARI of $0.87$ (median $=1.00$), with the central $80\%$ of values ranging from $0.58$ to $1.00$. The GPT-generated survey exhibits slightly higher average stability, with a mean ARI of $0.93$ (median $=0.92$) and a $10$th–$90$th percentile range of $0.80$ to $1.00$. While both surveys occasionally produce lower-agreement runs, as reflected in minimum ARI values, the predominance of high ARI scores indicates that the two-cluster solutions are robust to substantial perturbations of the data.
These results suggest that the dominant two-cluster structure recovered by both surveys reflects a stable latent division in climate beliefs rather than an artifact of sampling variability.

We next examine the degree to which the human baseline and GPT-generated surveys recover similar latent group structure. This analysis is important because high between-survey agreement indicates that both surveys are capturing a shared underlying belief organization, whereas low agreement may signal measurement distortion or instrument-specific artifacts.

The transition matrix in Table~\ref{tab:cluster-transition} shows strong agreement between the two surveys. Most respondents remain on the diagonal ($86\%$ and $98\%$), with minimal cross-cluster reassignment. This indicates that the GPT-generated survey recovers the same primary division in climate beliefs identified by the human baseline, with only limited redistributions.
\begin{table}[t]
\centering
\begin{tabular}{lcc}
\hline
\textbf{Human Baseline/GPT} & \textbf{Cluster 0} & \textbf{Cluster 1} \\
\hline
\textbf{Cluster 0} & 0.86 & 0.14 \\
\textbf{Cluster 1} & 0.02 & 0.98 \\
\hline
\end{tabular}
\caption{Transition matrix between clusters induced by the human baseline and the GPT-generated surveys. Rows correspond to human baseline clusters columns correspond to GPT clusters.}
\label{tab:cluster-transition}
\end{table}

We evaluate external semantic validity by examining how unsupervised cluster assignments align with persona labels that reflect respondents’ positions on climate change. As our primary analysis, we focus on alignment with self-identified personas, which were collected directly from participants and represent positions along a climate belief spectrum.

Using ARI and NMI, we observe modest but systematic alignment between clusters and self-identified personas. The human baseline survey achieves an ARI of $0.12$ and an NMI of $0.22$, while the GPT-generated survey achieves slightly higher alignment (ARI $=0.15$, NMI $=0.27$). 

Because persona labels include six categories while clustering yields only two groups, perfect alignment is neither expected nor feasible. The low ARI largely reflects this difference in granularity, whereas the non-trivial NMI indicates shared information beyond chance, suggesting both surveys capture a dominant ideological division underlying finer persona distinctions.

Inspection of cluster–persona distributions reveals a clear qualitative pattern. In both surveys, one cluster is enriched for self-identified personas expressing higher climate concern (e.g., \emph{Advocate for Proactive Climate Policy} and \emph{Climate Action Advocate}), while the other cluster is enriched for more skeptical or disengaged personas (e.g., \emph{Climate Policy Skeptic} and \emph{Advocate for Avoiding Climate Alarmism}), with neutral personas distributed between clusters. This structure suggests that unsupervised clustering recovers a meaningful ideological ordering.

Taken together, these results suggest that both survey instruments capture socially meaningful climate belief structure without trivially reproducing persona labels. Alignment increases as persona labels are coarsened, indicating that both surveys reliably recover a dominant ideological division, while differences between instruments emerge primarily in how much finer-grained nuance they preserve beyond this shared structure.

\subsection{Predictive Validity} \label{sec:results:prediction}

We evaluate predictive validity by assessing how well each survey predicts two downstream outcomes: support for policy and engagement in taking action. For each outcome, we first remove survey items that directly measure that outcome (e.g., policy-support questions when predicting policy support, action-oriented questions when predicting engagement) to prevent information leakage. The remaining survey questions are used as predictive features. In parallel, clusters are derived using only these non-outcome questions with similar clustering method as section \ref{sec:results:cluster}. For each outcome, we compare models trained on survey responses (feature-based models) to models that use only the induced cluster label as a predictor (cluster-based models). All models are estimated using Ridge regression, and predictive performance is measured using 5-fold cross-validated $R^2$.  We report results for all three topics since predictive performance varies by domain and modeling choice rather than favoring a single instrument uniformly( see Table~\ref{tab:prediction_results_by_topic}).

\begin{table}[t]
\centering
\small
\begin{tabularx}{\linewidth}{lXXXX}
\hline
\textbf{Survey / $R^2$} 
& \textbf{Policy (Feat.)}
& \textbf{Policy (Clust.)}
& \textbf{Action (Feat.)}
& \textbf{Action (Clust.)} \\
\hline

\multicolumn{5}{l}{\textbf{Climate Change}} \\
GPT & $0.486$ & $0.368$ & $0.158$ & $0.176$ \\
Human Baseline & $0.554$ & $0.426$ & $0.391$ & $0.144$ \\
\hline

\multicolumn{5}{l}{\textbf{Immigration}} \\
GPT  & $0.345$ & $0.154$ & $0.537$ & $0.488$ \\
Human Baseline & $0.285$ & $0.313$ & $0.525$ & $0.287$ \\
\hline

\multicolumn{5}{l}{\textbf{DEI}} \\
GPT  & $0.504$ & $0.450$ & $0.497$ & $0.421$ \\
Human Baseline  & $0.501$ & $0.592$ & $0.331$ & $0.244$ \\
\hline
\end{tabularx}
\caption{Predictive performance ($R^2$, mean 5-fold cross-validated values) using Ridge regression for feature-based (Feat.) and cluster-based (Clust.) models across all three topics.}
\label{tab:prediction_results_by_topic}
\end{table} 

For climate change, feature-based models using the human baseline survey outperform those using the GPT-generated survey for both policy support ($R^2=0.554$ vs.\ $0.486$) and action likelihood ($R^2=0.391$ vs.\ $0.158$). However, cluster-based performance narrows this gap, with GPT clusters slightly outperforming human clusters for action ($0.176$ vs.\ $0.144$), suggesting both surveys recover a similar coarse behavioral division.

For Immigration, the pattern shifts. GPT feature-based models slightly outperform the human baseline survey for policy prediction ($0.345$ vs.\ $0.285$), while action prediction is comparable at the feature level ($0.537$ vs.\ $0.525$). Notably, GPT clusters substantially outperform human baseline clusters for action prediction ($0.488$ vs.\ $0.287$), indicating stronger segmentation for this domain.

For DEI, feature-based policy prediction is nearly identical ($0.504$ vs.\ $0.501$), while GPT features yield higher predictive performance for action ($0.497$ vs.\ $0.331$). At the cluster level, the pattern reverses for policy: human baseline clusters outperform GPT clusters ($0.592$ vs.\ $0.450$), whereas GPT clusters perform better for action ($0.421$ vs.\ $0.244$), revealing domain-dependent differences in how belief structure translates to behavioral outcomes.

Overall, these results indicate that both survey types recover behaviorally relevant structure, but their predictive strengths vary by topic and by level of representation (item-level versus cluster-level). Rather than demonstrating uniform superiority of one instrument, the findings suggest domain-dependent differences in how belief structure translates into downstream behavioral prediction. Future work should focus on understanding the mechanisms underlying these differences. Identifying these sources of divergence may enable principled refinement of LLM-generated survey instruments and help close predictive gaps where they emerge.

\section{Discussion} 

Across three domains, GPT-generated surveys recover the dominant ideological structure identified by human baseline instruments. Clustering quality, stability, and cross-survey agreement indicate similar high-level respondent partitions. Differences arise primarily in how belief variation is represented beyond this dominant split. Dimensional efficiency and predictive analyses suggest that human baseline surveys often preserve a stronger fine-grained signal at the item level, while GPT-generated surveys can produce competitive segmentation in certain domains. Alignment with self-identified personas suggests that both surveys capture a shared ideological backbone underlying more nuanced distinctions. Further examination of this layered structure remains an important direction for future work.

These findings suggest that divergence between instruments lies less in the existence of structure and more in its resolution and translation into downstream prediction. Future work should examine how prompt design influences survey quality, evaluate generalization across additional topics, explore finer-grained clustering solutions, and investigate when predictive gaps are substantively meaningful.

Overall, LLM-generated surveys show strong promise for recovering coarse belief structure. However, the current evidence does not support their use as full replacements for expert-designed instruments without human oversight. In particular, our evaluation is limited to three social domains, and performance may vary for more technical, less salient, or more culturally specific topics. Instead, LLM-generated surveys may serve as useful scaffolds for exploratory analysis, pilot studies, or supervised survey development.

\section{Acknowledgements}

This work was partially supported by a Stony Brook University Climate Change Seed Grant. We gratefully acknowledge the Department of Political Science at Stony Brook University for providing funding for the crowdsourced user studies. This material is also based upon work supported by NSF under Grant NRT-HDR 2125295. Any opinions, findings and conclusions or recommendations expressed in this material are those of the author(s) and do not necessarily reflect the views of the National Science Foundation.

\section{Ethical Considerations}
This study involved human participants and was approved by the relevant Institutional Review Board (IRB). All participants provided informed consent prior to participation and were recruited through an online research platform. Participants were required to be 18 years of age or older and residing in the United States. No personally identifiable information was collected, and all analyses were conducted on anonymized data. Participants were compensated at a rate aligned with platform norms.

The survey topics examined (climate change, immigration, and DEI) are socially and politically sensitive. To minimize potential harm, survey questions were designed to be neutral, non-leading, and accessible to a general adult audience. Participants were free to discontinue participation at any time without penalty.

Because this study evaluates LLM-generated survey instruments, we also note the potential risks associated with deploying automated surveys in real-world settings. LLM-generated instruments may inadvertently reflect biases present in training data or oversimplify complex belief structures. Our findings do not advocate replacing expert-designed surveys with fully automated systems; rather, we emphasize the importance of human oversight when using LLMs in survey design and interpretation.

\section{Bibliographical References}
\bibliographystyle{lrec2026-natbib}
\bibliography{lrec2026-example}

\appendix
\section{Full GPT Prompts}
\label{app:full-prompts}

This appendix provides the complete prompt templates and generation parameters used to produce the GPT-generated surveys analyzed in this study. We include the full system and user prompts verbatim to support transparency, reproducibility, and careful interpretation of the survey generation process.

\begin{lstlisting}
MODEL = "gpt-4.1"
TEMPERATURE = 0.7
MAX_TOKENS = 800

SYSTEM_PROMPT = """
You are an expert survey methodologist and social scientist.
Your task is to design concise, signal-rich surveys suitable for
persona discovery and cluster analysis.
"""

USER_PROMPT_TEMPLATE = """
I want to understand how people think, feel, and act with respect to the following topic:

Topic: "{topic}"

Design a concise, signal-rich survey that captures the major latent
dimensions of human stance on this topic.

Survey requirements:
1. The survey must contain exactly 9 questions arranged as a 3 by 3 structure:
   - 3 belief / cognitive questions
   - 3 perception / affect questions
   - 3 behavior / intention / policy questions

2. Each question must:
   - Be answerable on a 5-point Likert scale
   - Be neutral and non-leading
   - Be understandable to a general adult audience
   - Avoid technical jargon

3. The questions should be:
   - Non-redundant
   - Designed to maximize differentiation between respondents
   - Capable of revealing distinct personas when analyzed jointly

Output format:
- Group questions into three sections (Beliefs, Perceptions, Behaviors)
- Number questions from 1 to 9
- For each question, include a brief parenthetical note describing
  the latent dimension it probes

Only output the survey. Do not include analysis or commentary.
"""
\end{lstlisting}

\section{Full Survey Questions and Self-Identified Labels}
\subsection{Climate Change}
\subsubsection{Persona Labels and Descriptions}
The following self-identification question was presented to the users after signing the consent form.

\textbf{Which of the following labels and descriptions best reflect your view on climate change?}
\begin{itemize}
    \item  \textbf{Advocate for Avoiding Climate Alarmism}\\
    You are a person who does not believe that climate change is a real or serious issue, particularly not one caused by human activity. You believe climate-related policies are unnecessary and may harm economic growth or personal freedoms. You generally oppose environmental regulations and favor deregulation.
    \item \textbf{Climate Policy Skeptic}\\
    You are a person who believes climate change is happening but questions the degree of threat and the effectiveness of government-led solutions. You may worry that some climate policies go too far or could harm economic opportunity and personal freedoms.
    \item \textbf{Neutral observer of Climate Change}\\
    You are someone who recognizes that environmental concerns exist but doesn’t follow climate debates closely. You may adopt eco-friendly practices when convenient, but climate action is not a central concern in your daily life.
    \item \textbf{Advocate for Balanced Climate Action}\\
    You are a person who believes climate change is real and worth addressing, but solutions should be practical and economically responsible. You support action that balances environmental goals with innovation, affordability, and individual freedom.
    \item \textbf{Advocate for Proactive Climate Policy}\\
    You believe climate change is real and requires action, but think solutions should balance environmental responsibility with economic and social considerations. You support investments in renewable energy, resilience, and policy reform that is both effective and sustainable.
    \item \textbf{Climate Action Advocate}\\
    You are someone who sees climate change as an urgent challenge requiring bold action. You support ambitious policy reforms, investment in renewables, and efforts to hold corporations accountable. You may also adopt lifestyle changes to reduce your personal impact and promote environmental responsibility.
\end{itemize}

\subsubsection{Human Baseline Survey}
The human baseline survey consisted of items drawn from established instruments. The first four questions were adapted from SASSY \cite{chryst2018global}, and Question 5 was taken from the Climate Change Perception Questionnaire \cite{poortinga2019climate}. These were followed by items V241749 and V242321 from the ANES survey \cite{anes2020ts}, as well as a single-item measure of belief in climate change \cite{berger2025measuring}. The resulting survey items are listed below. Question 6 was one of the questions used in section \ref{sec:results:prediction} for predicting policy support. 

\begin{enumerate}
    \item How important is the issue of climate change to you personally?\\
    1. Not at all important, 2. a little important, 3. moderately important, 4. very important, 5. extremely important
    \item How worried are you about global warming?\\
Very worried 2. Somewhat worried 3. Not very worried 4. Not at all worried
    \item How much do you think global warming will harm you personally?\\
1.A great deal  2. A moderate amount 3. Only a little 4. Not at all
    \item How much do you think global warming will harm future generations of people?\\
    1. not at all, 2. only a little, 3. a moderate amount, 4. a great deal
    \item How good or bad do you think the impact of climate change would be on people across the world?\\
    1. Extremely bad, 2. Bad, 3.Neither good or bad, 4. Good, 5.Extremely good
    \item Do you favor, oppose, or neither favor nor oppose increased government regulation on businesses that produce a great deal of greenhouse emissions linked to climate change?\\
    1. Favor, 2. Oppose, 3. Neither favor nor oppose\\\\
For each of the following statements, please indicate the extent to which you agree or disagree using the following scale: 1. Strongly Agree 2. Somewhat Agree 3. Neutral 4. Somewhat Disagree 5. Strongly Disagree.
    \item  Climate change is currently affecting severe weather events or temperature patterns in the United States.
    \item The occurrence of climate change is caused by human activities and will bring largely negative consequences.

\end{enumerate}

\subsubsection{GPT Survey}
 The resulting survey items from the prompt provided in section \ref{app:full-prompts} are listed below. Questions 7 and 8 were questions used in section \ref{sec:results:prediction} for predicting policy support and downstream action. 
\\

For each statement below, please indicate the extent to which you agree or disagree. Please respond honestly. There are no right or wrong answers.

Response scale (for the next 6 items below): 
A great deal; A lot; A moderate amount; A little; None at all.

\begin{enumerate}
    \item How much do you agree that climate change is happening?
    \item To what extent do you think human activities contribute to climate change?
    \item How confident are you in the scientific information available about climate change?
    \item How concerned do you feel about the effects of climate change on the planet?
    \item How serious do you believe the consequences of climate change are for people in your country?
    \item How hopeful do you feel about society’s ability to address climate change?

\item How often do you take actions to reduce your own environmental impact (such as conserving energy, reducing waste, or choosing sustainable products)?

 1. Always 2. Often 3. Sometimes 4. Rarely 5.Never.

 \item How likely are you to support policies or laws aimed at reducing climate change, even if they have economic costs?

1. Extremely likely 2. Very likely 3. Moderately likely 4. Slightly likely 5. Not at all likely.

\item How likely are you to discuss climate change or related topics with others in your social circle?

1. Extremely likely 2. Very likely 3. Moderately likely 4.Slightly likely 5. Not at all likely.

\end{enumerate}

\subsection{Immigration}
\subsubsection{Persona Labels and Descriptions}
The following self-identification question was presented to the users after signing the consent form.
\textbf{Which of the following labels and descriptions best reflect your view on Immigration?}
\begin{itemize}
    \item \textbf{Advocate for Border Integrity}\\
You prioritize secure borders, the enforcement of immigration laws, and the preservation of national identity. You believe immigration should be tightly controlled.
    \item \textbf{Advocate for Selective Immigration}\\
You support legal immigration that serves the country’s strategic goals. You favor policies that emphasize skills, economic contribution, or national security, while being cautious about broad or open-ended immigration reforms.
    \item \textbf{Neutral Observer}\\
You recognize immigration as a complex issue and prefer not to take a fixed stance. You believe immigration policies should be shaped by context, evidence, and evolving societal needs rather than ideology.
    \item \textbf{Advocate for Balanced Immigration}\\
You support immigration policies that balance national interests with humanitarian values. You favor fair and efficient systems that welcome newcomers while maintaining social cohesion and economic stability.
    \item \textbf{Pragmatic Immigration Supporter}
You support humane and efficient immigration processes that weigh both national interests and human dignity. You believe reforms should be practical and inclusive, ensuring economic contribution while offering protection to those in need.
    \item \textbf{Advocate for Global Mobility}
You are a person who strongly supports welcoming immigration policies that offer opportunities to those seeking a better life. You advocate for inclusive, well-resourced systems that help immigrants integrate into society and contribute fully.

\end{itemize}
\subsubsection{Human Baseline Survey}
For the immigration domain, we relied on items drawn from ANES \cite{anes2020ts}, which includes a broad set of questions capturing multiple aspects of public attitudes toward immigration, including perceived impacts, fairness, and policy preferences. From this item pool, we selected nine non-redundant questions (specifically ANES items V201424, V201427, V202240, V202419, V202246, V202420, V202418, V201417, and V201421) to construct a human survey baseline with coverage across belief, perception, and behavior-oriented dimensions. This selection allowed us to preserve the diversity of immigration-related attitudes represented in ANES while matching the length and overall structure of the GPT-generated surveys. The resulting survey items are listed below. Questions 8 and 9 were the questions used in section \ref{sec:results:prediction} for predicting policy support.

For each statement below, please indicate the extent to which you agree or disagree. Please respond honestly. There are no right or wrong answers.\\
Response scale (for the next seven items below): 
Strongly Agree; Somewhat Agree; Neutral; Somewhat Disagree; Strongly Disagree.
\begin{enumerate}
    \item A wall should be built at the U.S. border with Mexico.
    \item It is important that everyone in the United States learn to speak English.
    \item There should be a path to citizenship for unauthorized immigrants who obey the law, pay a fine, and pass security checks.
    \item America's culture is generally harmed by immigrants.
    \item The children of detained immigrants should be separated rather than kept with their parents in adult detention centers.
    \item Immigrants increase crime rates in the United States.
    \item Immigrants are generally good for America's economy.
    \item Which comes closest to your view about what government policy should be toward unauthorized immigrants now living in the United States?
    \begin{itemize}
    \item Make all unauthorized immigrants felons and send them back to their home country.
    \item Have a guest worker program that allows unauthorized immigrants to remain in the United States in order to work, but only for a limited amount of time.
    \item Allow unauthorized immigrants to remain in the United States and eventually qualify for U.S. citizenship, but only if they meet certain requirements like paying back taxes and fines, learning English, and passing background checks.
    \item Allow unauthorized immigrants to remain in the United States and eventually qualify for U.S. citizenship, without penalties.
\end{itemize}
    \item What should happen to immigrants who were brought to the U.S. illegally as children and have lived here for at least 10 years and graduated high school here?
    \begin{itemize}
    \item Should be sent back where they came from.
    \item Should be allowed to live and work in the U.S.
\end{itemize}

\end{enumerate}

\subsubsection{GPT Survey}
 The resulting survey items from the prompt provided in section \ref{app:full-prompts} are listed below. Questions 7,8, and 9 were questions used in section \ref{sec:results:prediction} for predicting policy support and downstream action.

 For each statement below, please indicate the extent to which you agree or disagree. Please respond honestly. There are no right or wrong answers.
Response scale (for all items below): 
Strongly Agree; Somewhat Agree; Neutral; Somewhat Disagree; Strongly Disagree.
\begin{enumerate}
    \item Immigration has a positive effect on the economy of the receiving country.
    \item The culture of this country is strengthened by the presence of people from different backgrounds.
    \item The number of immigrants entering my country in recent years is higher than is good for society.
    \item I feel comfortable interacting with people who have recently moved here from other countries.
    \item News stories about immigration often make me feel anxious or concerned.
    \item I worry that immigration could make it harder for me or my family to find opportunities.
    \item I would support policies that make it easier for people from other countries to live and work here.
    \item I would be willing to sign a petition supporting the rights of immigrants.
    \item I believe the government should prioritize strict enforcement of immigration laws.
\end{enumerate}
\subsection{DEI}
\subsubsection{Persona Labels and Descriptions}
The following self-identification question was presented to the users after signing the consent form.
\textbf{Which of the following labels and descriptions best reflect your view on DEI?}

DEI stands for Diversity, Equity, and Inclusion: principles aimed at promoting fair treatment and meaningful participation for all people, especially those historically marginalized or underrepresented.
\begin{itemize}
    \item \textbf{Advocate for Meritocracy}\\
You are someone who believes that fairness means treating everyone the same, regardless of identity. You are cautious about DEI initiatives that prioritize group-based outcomes, as you worry they may compromise neutrality or create unintended divisions. You prefer merit-based systems and evaluate policies based on their effectiveness and broad impact.
\item \textbf{Advocate for Individual-Centered Fairness}
You value fairness on an individual level, believing that opportunities should be tailored to people’s qualifications and efforts rather than group membership. You may support inclusive efforts when they focus on shared human dignity and not identity-based metrics.
\item \textbf{Neutral Observer}
You are someone who hasn’t thought much about DEI and doesn’t feel strongly either way. You believe in treating everyone fairly, but DEI isn’t something you follow closely or prioritize. You tend to focus on your day‑to‑day responsibilities and judge DEI efforts based on what you experience locally.
\item \textbf{Advocate for Balanced Opportunity}
You support efforts to promote fairness and inclusion, but believe they should be measured and context-sensitive. You favor approaches that expand access without compromising merit, cohesion, or organizational effectiveness.
\item \textbf{Advocate for Inclusive Culture}
You are someone who believes that creating a respectful, inclusive culture helps everyone thrive. You support structured DEI efforts—like internal training, inclusive practices, and broader representation—as essential for fostering belonging, engagement, and team success.
\item \textbf{Advocate for Structural Equity}
You are someone who is deeply committed to systemic equity. You work to dismantle institutional barriers and amplify marginalized voices. You advocate for transparent policies, structural reform, and sustained investment to ensure long-term, accountable change.
\end{itemize}

\subsubsection{Human Baseline Survey}
For the DEI domain, we reviewed several established survey instruments measuring attitudes toward affirmative action, diversity, merit, and discrimination, including debate-style items from large-scale surveys such as the General Social Survey (GSS) \cite{gss2022} and ANES \cite{anes2020ts}, as well as multi-item scales from the social psychology literature. While GSS and ANES items are widely used, they primarily capture overall support or opposition through single questions with limited construct differentiation.

For the human survey baseline, we therefore drew on the Diversity, Merit, Fairness, and Discrimination (DMFD) Belief Scales \cite{aberson2007diversity}, which decompose DEI-related attitudes into multiple interpretable constructs using parallel item structures. To ensure comparability with the GPT-generated surveys, we selected three out of four constructs, resulting in a nine-item instrument matched in length and overall structure to the GPT-generated 3×3 surveys. The resulting survey items are listed below. Questions 1, 2, and 3 were the questions used in section \ref{sec:results:prediction} for predicting policy support.

For each statement, please indicate the extent to which you agree or disagree. Please respond honestly. There are no right or wrong answers.

\textit{Affirmative action refers to policies or practices designed to increase opportunities for historically underrepresented groups in areas such as education and employment. These measures may include considering factors like race, ethnicity, or gender as part of admissions or hiring decisions, with the goal of promoting diversity and addressing past discrimination.}

Response scale (for all items below): 
Strongly Agree; Somewhat Agree; Neutral; Somewhat Disagree; Strongly Disagree.

\begin{enumerate}
    \item Affirmative-action hiring policies are fair.
    \item Affirmative-action policies give everyone an equal chance.
    \item Affirmative-action policies are unfair to White men.
    \item Only the most qualified applicant should be hired, regardless of race or gender.
    \item Hiring decisions should be based solely on merit.
    \item Considering race or gender in hiring violates the principle of merit.
    \item A diverse workforce benefits an organization.
    \item Employees from different backgrounds improve problem solving.
    \item Diversity makes companies stronger.
    \item Racial discrimination in hiring is still common today.
    \item Minorities do not yet have equal job opportunities.
    \item Many employers would refuse a qualified minority applicant.
\end{enumerate}

\subsubsection{GPT Survey}
The resulting survey items from the prompt provided in section \ref{app:full-prompts} are listed below. Questions 7,8, and 9 were questions used in section \ref{sec:results:prediction} for predicting policy support and downstream action. 

For each statement below, please indicate the extent to which you agree or disagree. Please respond honestly. There are no right or wrong answers.

Response scale (for all items below): 
Strongly Agree; Somewhat Agree; Neutral; Somewhat Disagree; Strongly Disagree.

\begin{enumerate}
    \item Having people from different backgrounds improves group performance.
    \item Efforts to promote fairness and inclusion are necessary to address existing inequalities.
    \item Opportunities should be based solely on individual merit, regardless of background or identity.
    \item I feel comfortable interacting with people whose backgrounds or perspectives are different from my own.
    \item I perceive that people like me are treated fairly in most settings (e.g., work, school, public spaces).
    \item I often notice unfair treatment or bias based on identity (such as race, gender, ability, etc.) in my everyday life.
    \item I am likely to support policies or programs that aim to increase representation of underrepresented groups.
    \item I am likely to speak up or take action if I witness discrimination or unfair treatment.
    \item I actively seek out information or experiences that expose me to different perspectives.
\end{enumerate}
\section{User Demographic Information}

We recruited participants online for each topic-specific survey. After data cleaning and attention checks, the final sample consisted of 144 respondents for the climate change survey, 170 for the immigration survey, and 167 for the DEI survey. Participants were restricted to U.S.-based respondents and varied across gender, race/ethnicity, education level, and age group. The demographic distribution for each topic is summarized in Table \ref{tab:demographics}.



\begin{table}[h]
\centering
\small
\resizebox{\linewidth}{!}{%
\begin{tabular}{lccc}
\hline
\textbf{Demographic (\%)} & \textbf{Climate} & \textbf{Immigration} & \textbf{DEI} \\
\hline
\multicolumn{4}{l}{\textit{\textbf{Gender}}} \\
Female &  $50.7$& $44.7$ & $52.1$\\
Male & $48.6$ & $52.9$ & $46.1$\\
Other & $0.7$ & $2.4$ & $1.8$\\

\multicolumn{4}{l}{\textit{\textbf{Race / Ethnicity}}} \\
White & $72.2$ & $78.2$& $72.5$\\
Black or African American & $11.8$ & $10$& $13.8$ \\
Asian & $8.3$ & $4.7$ & $6.6$\\
Other / Multiracial & $7.7$ & $7.1$ & $7.1$\\

\multicolumn{4}{l}{\textit{\textbf{Education}}} \\
High school or less & $15.2$ & $17.1$ & $13.4$\\
Some college & $30.6$ &$34.7$ & $30.6$\\
Bachelor’s degree & $38.9$ & $30.6$ & $35$\\
Graduate degree &$15.3$  & $17.6$ & $21$\\

\multicolumn{4}{l}{\textit{\textbf{Age Group}}} \\
18–24 & $9$ & $11.2$& $7.8$ \\
25–34 & $24.3$ & $26.5$ & $29.9$\\
35–44 & $29.2$ &$32.4$ & $28.7$ \\
45-54 &$13.9$ & $14.1$& $12$\\
55+ & $23.6$ & $15.8$ & $21.6$\\
\hline
\end{tabular}
}
\caption{Participant demographic characteristics (Climate: $N=144$; Immigration: $N=170$; DEI: $N=167$).}
\label{tab:demographics}
\end{table}
\section{Results}
\subsection{Immigration}

\begin{figure}[t]
    \centering
    \includegraphics[width=\linewidth]{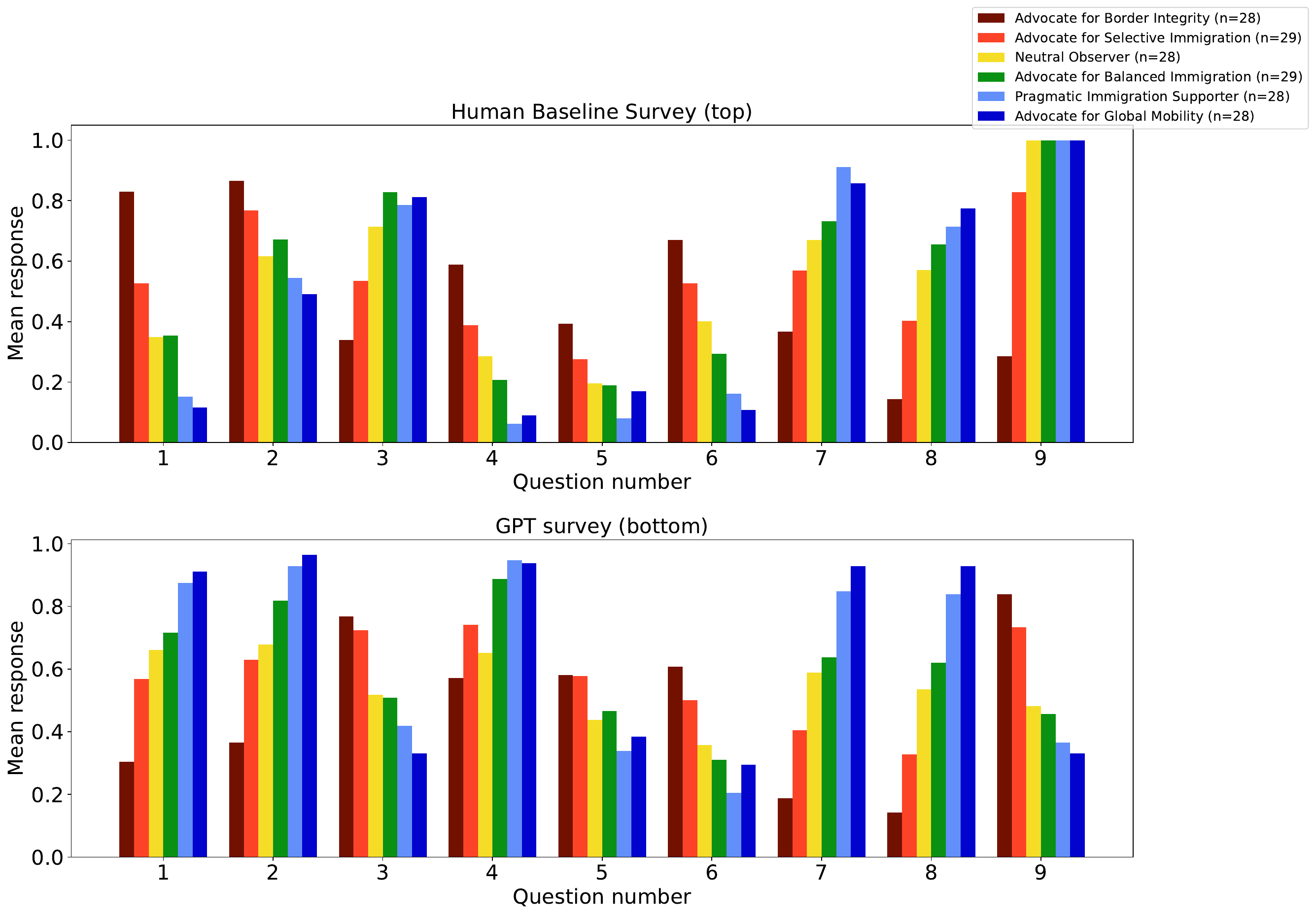}
    \caption{
    Immigration surveys: Normalized mean response per question by persona label.
    The top panel shows the human baseline survey and the bottom panel shows the GPT-generated survey.
    Bars are colored by self-identified immigration persona, ordered from least to most concerned.}
    \label{fig:means_by_persona_immigration}
\end{figure}

Figure~\ref{fig:means_by_persona_immigration} presents the mean response for each question in the human baseline survey (top) and the GPT-generated survey (bottom), with bars colored by immigration persona. Across most questions in both surveys, responses exhibit a clear and ordinal pattern: as one moves along the persona spectrum( from \textit{Advocate for Border Integrity} to \textit{Advocate for Global Mobility}) mean responses shift systematically in directions consistent with increasing support for immigration.

This progression indicates that both instruments capture a meaningful underlying gradient of immigration attitudes rather than purely idiosyncratic variation. In the human baseline survey, persona means generally follow the expected ordering across items, with relatively clear separation between opposing ends of the spectrum, though spacing between adjacent personas varies by question. The GPT-generated survey reproduces the broad ordinal structure across most items, but the distinctions between neighboring personas are more compressed for several questions, and separation is less consistent across the full set of items.

Overall, the results suggest that self-identified immigration personas correspond to systematic response differences, and that both the human-designed and GPT-generated surveys reflect these distinctions in an interpretable, though not perfectly uniform, manner.

\subsubsection{Clustering Quality and Stability}

\begin{figure}[t]
    \centering

    \begin{subfigure}{\linewidth}
        \centering
        \includegraphics[width=\linewidth]{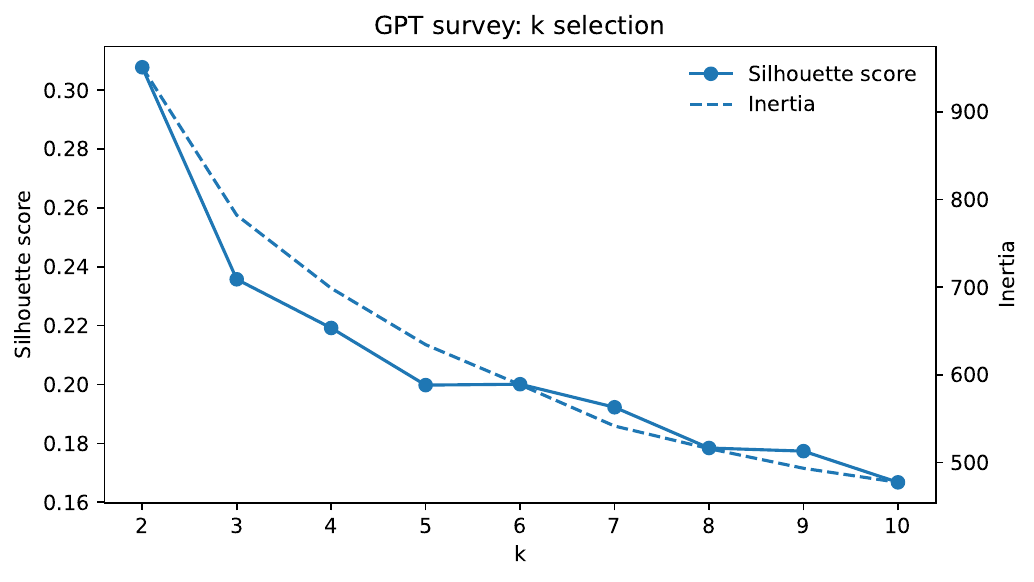}
    \end{subfigure}

    \begin{subfigure}{\linewidth}
        \centering
        \includegraphics[width=\linewidth]{./figs/kmeans_k_selection_gpt_immigration.pdf}
    \end{subfigure}
    \caption{
    Selection of the number of clusters ($k$) using silhouette score and inertia for the human baseline (top) and GPT-generated (bottom) immigration change surveys. 
    }
    \label{fig:k_selection_immigration}
\end{figure}

We determine the cluster structure supported by each instrument using internal validation criteria. Both the Silhouette Score and the DBI indicate that a two-cluster solution ($k=2$) provides the best balance of cohesion and separation for the human baseline and GPT-generated immigration surveys (see the Silhouette curve in Figure~\ref{fig:k_selection_immigration}). Although DBI curves are omitted for brevity, they consistently favor the same solution. Accordingly, we proceed with $k=2$ in order to facilitate direct comparison between instruments.

To evaluate the robustness of this partition, we conduct a resampling-based stability analysis. For each survey, we perform $300$ bootstrap iterations, sampling $80\%$ of respondents without replacement and reclustering using $k$-means with $k=2$. Stability is quantified via the ARI, which measures agreement between cluster assignments across runs.

Both surveys yield highly stable partitions. The human baseline produces a mean ARI of $0.79$ (median $=0.835$), with the central $80\%$ of values spanning $0.53$ to $1.00$. The GPT-generated survey exhibits slightly higher average stability, with a mean ARI of $0.82$ (median $=0.86$) and a $10$th--$90$th percentile range of $0.58$ to $1.00$. Although a small number of runs show reduced agreement, the predominance of high ARI values indicates that the two-cluster structure is robust to substantial perturbations of the data.

We next evaluate the extent to which the human baseline and GPT-generated surveys yield comparable latent partitions of respondents. The transition matrix in Table~\ref{tab:cluster_transition_immigration} indicates substantial concordance between the two surveys. A large majority of respondents remain on the diagonal ($76\%$ and $95\%$), with relatively few reassigned across clusters. This pattern suggests that the GPT-generated survey recovers the same dominant attitudinal division identified by the human baseline, with only modest reallocation near the boundaries of the partition.

\begin{table}[t]
\centering
\begin{tabular}{lcc}
\hline
\textbf{Human Baseline/GPT} & \textbf{Cluster 0} & \textbf{Cluster 1} \\
\hline
\textbf{Cluster 0} & 0.76 & 0.24 \\
\textbf{Cluster 1} & 0.05 & 0.95 \\
\hline
\end{tabular}
\caption{Transition matrix between clusters induced by the human baseline survey and the GPT-generated survey. Rows correspond to human baseline clusters; columns correspond to GPT clusters.}
\label{tab:cluster_transition_immigration}
\end{table}

We assess external semantic validity by examining how unsupervised cluster assignments align with participants’ self-identified personas, which were collected at the beginning of the study and reflect ordered positions along a immigration belief spectrum. This comparison evaluates whether the latent groupings induced by each survey correspond to respondents’ own stated ideological orientations.

Using the ARI and NMI, we find modest but systematic alignment between clusters and persona labels. The human baseline survey yields an ARI of $0.12$ and an NMI of $0.21$, while the GPT-generated survey shows slightly higher alignment (ARI $=0.15$, NMI $=0.22$). Although these values do not indicate one-to-one correspondence, they reflect non-trivial shared structure between unsupervised clusters and self-reported identities.

\subsection{DEI}

\begin{figure}[t]
    \centering
    \includegraphics[width=\linewidth]{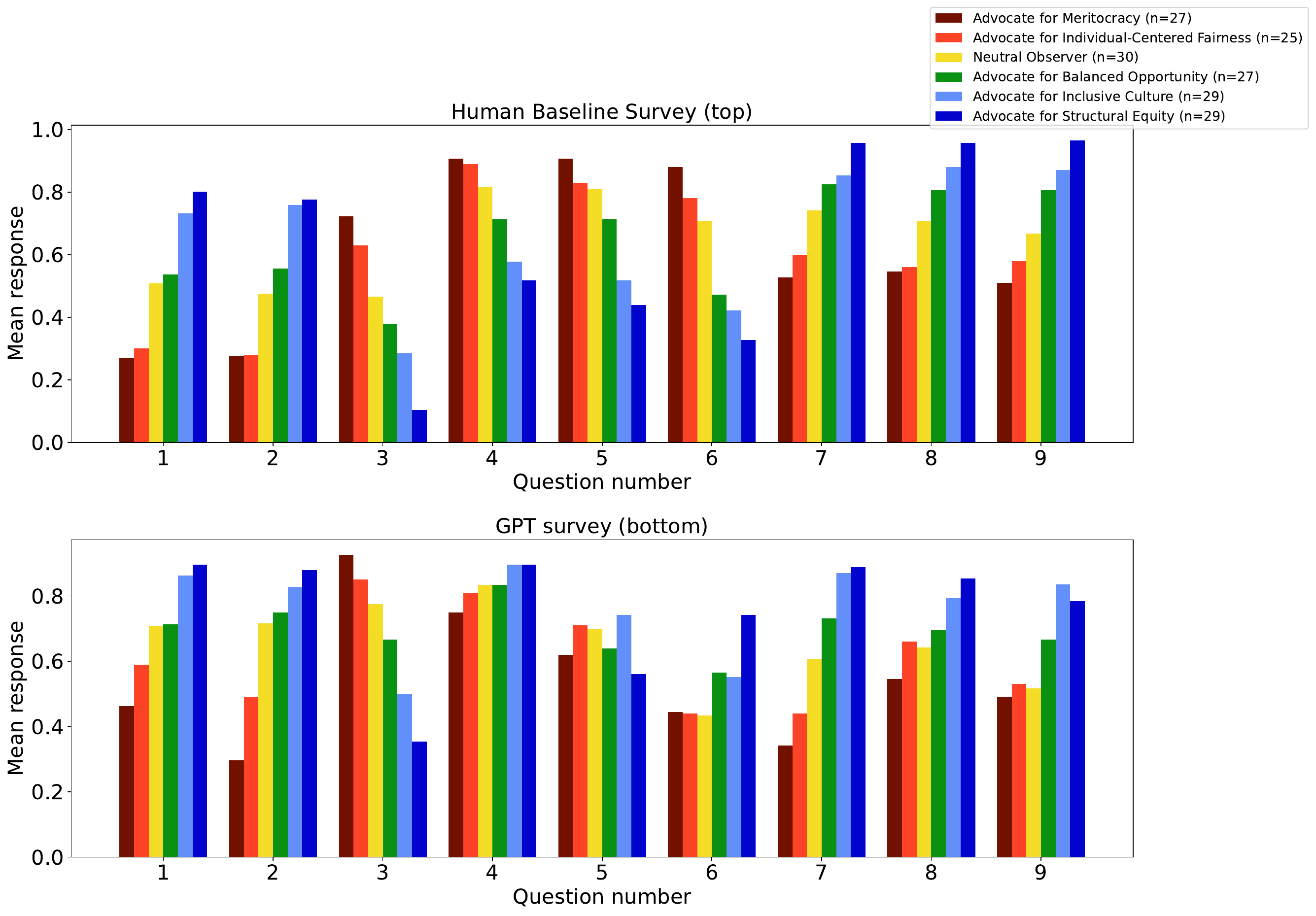}
    \caption{
    DEI surveys: Normalized mean response per question by persona label.
    The top panel shows the human baseline survey and the bottom panel shows the GPT-generated survey.
    Bars are colored by self-identified DEI persona, ordered from least to most concerned.}
    \label{fig:means_by_persona_DEI}
\end{figure}

Figure~\ref{fig:means_by_persona_DEI} presents the mean response for each question in the human baseline survey (top) and the GPT-generated survey (bottom), with bars colored by DEI persona. 

In the DEI domain, the persona patterns are highly structured and largely mirrored across the human baseline (top) and GPT survey (bottom). For several items (especially Questions 1–2 and 7–9) mean responses increase steadily from \textit{Advocate for Meritocracy} through to \textit{Advocate for Structural Equity}, with the highest-endorsement personas consistently at the inclusive-culture/structural-equity end of the spectrum. For other items (most clearly Questions 3–6 in the human baseline), the direction flips: meritocracy-oriented personas show higher agreement and structural-equity personas lower agreement, which is exactly what we would expect for items framed around merit or skepticism toward identity-based interventions. The GPT-generated survey preserves this overall structure—strong separation at the extremes and coherent ordering on most items—though a few questions show small non-monotonicities (e.g., Q5 and Q9) where adjacent personas overlap or the top-end persona is not strictly the maximum. Overall, the figure suggests that both instruments capture the same underlying DEI stance spectrum, with item-dependent polarity reflecting whether an item aligns with meritocratic versus equity-centered framing.

\subsubsection{Clustering Quality and Stability}

\begin{figure}[t]
    \centering

    \begin{subfigure}{\linewidth}
        \centering
        \includegraphics[width=\linewidth]{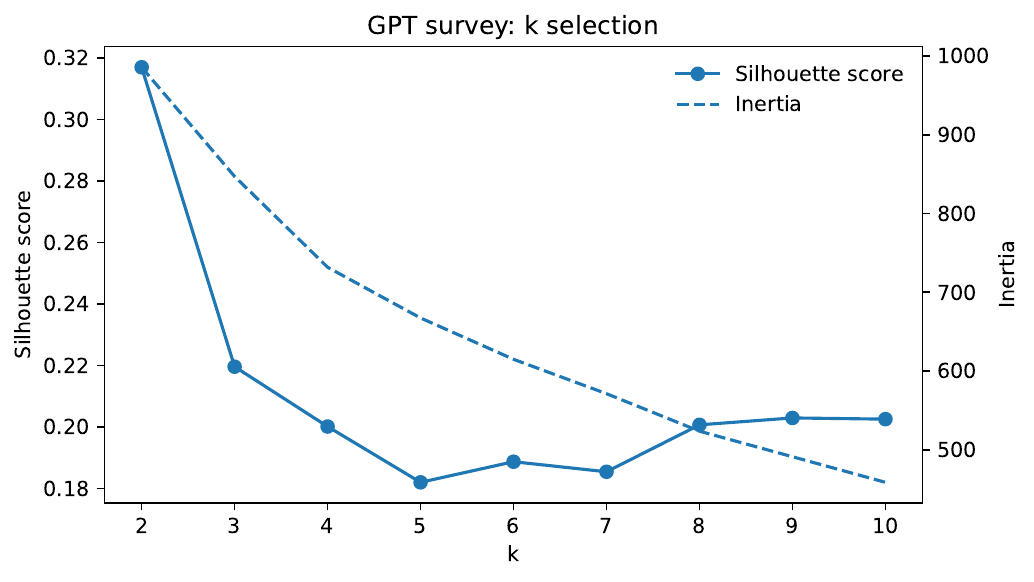}
    \end{subfigure}

    \begin{subfigure}{\linewidth}
        \centering
        \includegraphics[width=\linewidth]{./figs/kmeans_k_selection_gpt_DEI.pdf}
    \end{subfigure}
    \caption{
    Selection of the number of clusters ($k$) using silhouette score and inertia for the human baseline (top) and GPT-generated (bottom) DEI change surveys. 
    }
    \label{fig:k_selection_DEI}
\end{figure}

We next identify the cluster configuration favored by each survey using internal validation metrics. Both the Silhouette Score and the DBI converge on a two-cluster solution ($k=2$) as offering the strongest trade-off between within-cluster cohesion and between-cluster separation for the human baseline and GPT-generated DEI surveys (see the Silhouette curve in Figure~\ref{fig:k_selection_DEI}). We therefore adopt $k=2$ for subsequent analyses to enable direct comparison across instruments.

As in the other domains, we evaluate the stability of the $k=2$ solution using the same resampling-based procedure described for other topics. Both instruments exhibit strong stability under resampling. The human baseline achieves a mean ARI of $0.86$ (median $=0.92$), with the central $80\%$ of values ranging from $0.63$ to $1.00$. The GPT-generated survey demonstrates comparable robustness, with a mean ARI of $0.83$ (median $=0.85$) and a $10$th--$90$th percentile interval of $0.60$ to $1.00$. While a small number of iterations yield lower agreement, the concentration of high ARI values indicates that the two-cluster structure is resilient to substantial perturbations of the sample.

We examine cross-instrument consistency by comparing the partitions produced by the human baseline and GPT-generated surveys. The transition matrix in Table~\ref{tab:cluster_transition_DEI} reveals strong alignment between the two solutions. A substantial proportion of respondents remain on the diagonal ($89\%$ and $74\%$), with relatively limited reassignment across clusters. This pattern suggests that the GPT-generated survey recovers the same primary attitudinal division identified by the human baseline, with differences largely confined to respondents near the decision boundary between clusters.

\begin{table}[t]
\centering
\begin{tabular}{lcc}
\hline
\textbf{Human Baseline/GPT} & \textbf{Cluster 0} & \textbf{Cluster 1} \\
\hline
\textbf{Cluster 0} & 0.89 & 0.11 \\
\textbf{Cluster 1} & 0.26 & 0.74 \\
\hline
\end{tabular}
\caption{Transition matrix between clusters induced by the human baseline survey and the GPT-generated survey. Rows correspond to human baseline clusters; columns correspond to GPT clusters.}
\label{tab:cluster_transition_DEI}
\end{table}

We assess external semantic validity following the same procedure used for the other two topics. Using the ARI and NMI, we observe modest but systematic alignment between clusters and persona labels. The human baseline survey yields an ARI of $0.07$ and an NMI of $0.12$, while the GPT-generated survey shows comparable alignment (ARI $=0.07$, NMI $=0.12$). Although these values are lower than those observed in the climate and immigration domains, they are similar across both instruments, suggesting that each survey captures a comparable degree of shared structure with self-reported DEI personas.

\end{document}